\documentclass[aps,twocolumn]{revtex4}%
\usepackage{amsfonts}
\usepackage{amsmath}
\usepackage{amssymb}
\usepackage{graphicx}
\usepackage{epstopdf}
\providecommand{\U}[1]{\protect\rule{.1in}{.1in}}

\begin{document}
\title[Short title for running header]{Scaling of Non-Saturating MR and quantum oscillations in pristine and ion-implanted HOPG}
\author{Nicholas Cornell}
\affiliation{University of Texas at Dallas, Richardson, TX}
\author{M. B. Salamon}
\affiliation{University of Texas at Dallas, Richardson, TX}
\author{A. Zakhidov}
\affiliation{University of Texas at Dallas, Richardson, TX}
\pacs{PACS number}

\begin{abstract}
A wide variety of resistive and field dependent behaviors have been previously observed in both doped and non-doped Highly Oriented Pyrolytic Graphite (HOPG). We find HOPG samples to vary significantly in their temperature dependent resistances, even between portions taken from the same sample, yet they exhibit consistent non-saturating magnetoresistance (MR).
 The scaling behavior of the MR is shown to be characteristic of a model based on the Hall effect in granular materials.  In addition to the large, field-linear MR, all samples exhibit Shubnikov-de Haas (SdH) oscillations.  Additional samples were doped via ion-implantation by boron and phosphorous, but show no signs of superconductivity nor any systematic change in their magnetoresistive behavior. Analysis of the SdH data gives a 2D carrier density in agreement with previous results, and a large mean-free path relative to crystallite size, even in samples with thin ion-implanted surface layers.

\end{abstract}
\volumeyear{year}
\volumenumber{number}
\issuenumber{number}
\eid{identifier}
\date[Date text]{date}
\maketitle

Highly oriented pyrolytic graphite (HOPG) is a commercial product that finds
uses in areas diverse as neutron monochrometers\cite{dachs} and ultra-flat
substrates for scanning probe microscopy.\cite{baro} More recently, hints of
magnetism and superconductivity have been reported in HOPG, the latter
reportedly extending to 400 K\cite{Kopelevich2000}. Ion implantation of HOPG with phosphorous has been reported to induce superconductivity at low doses of low energy ions.  Very large, non-saturating
magnetoresistance and evidence for Shubnikov-de Haas (SdH) oscillations have
also been found.\cite{Kopelevich1999,amdt} Large variations in transport and
magnetic properties between samples, and even on portions of the same sample,
are common.\cite{kempa} HOPG materials are typically characterized by their
degree of mosaicity, by which is meant the angular spread of the c-axis
orientation of component graphite crystallites. For example, ZYA quality is
stated to have a mosaic spread of $0.4^{\circ},$ and is of the highest quality
commercially available. Crystallite size (up to 10 $\mu$m for ZYA),
intergranular connectivity, and possible interstitial impurities are clearly
candidates for the large variation in electrical and magnetic properties.

\begin{figure}
[ptb]
\begin{center}
\includegraphics[
height=2.6455in,
width=3.4186in
]%
{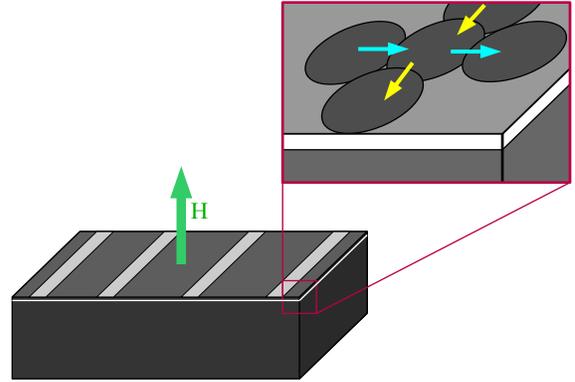}%
\caption{Schematic structure of the typical sample of HOPG of ZYA quality used in experiments, with four contact leads attached to the surface, and the thin ion-implanted surface layer shown. The inset shows the orbits of electrons crossing the interfaces between crystallites perpendicular to the direction of the electric field.}%
\label{FigM}%
\end{center}
\end{figure}

In this paper, we focus on magnetoresistive (MR) effects and attempt to
separate intrinsic properties of the component graphite crystallites from the
extrinsic effects associated with the composite nature of HOPG. We show that
the microscopic properties, as revealed by the SdH oscillations, are
independent of the bulk dc resistance. However, the very inhomogeneity of the
material, composed of crystallites of different sizes and orientation, leads to important physical effects, with nonsaturating
magnetoresistance\cite{Kopelevich2003} prominent among them. Similar
observations of linear magnetoresistance have been reported on silver
chalcogenides and were shown to persist to 60 T.\cite{hussman} An explanation
was proposed by Parish and Littlewood,\cite{parrish} who modeled the
inhomogeneous conductor as an array of coupled four-terminal circuit elements, as disks in which four currents can enter from adjacent elements.  Each disk is characterized by its scalar resistivity $\rho$ and Hall resistivity
$\rho\mu H$, where $\mu$ is the mobility and $H$, the applied field. As the
field increases, the Hall effect in each crystallite causes current to flow
through terminals that are normal to the electric field, giving rise to the
large linear MR that is observed. Allowing for an average mobility
$\left\langle \mu\right\rangle $ and a standard deviation $\Delta\mu$, Parish and
Littlewood determined the crossover between quadratic dependence on magnetic
field, for fields below a characteristic value $H_{0}$, and linear behavior
above. For $\Delta\mu/\left\langle \mu\right\rangle $ small, as we might
expect for HOPG, the crossover field is of order $1/\left\langle
\mu\right\rangle $ and the overall scale of the MR, proportional to
$\left\langle \mu\right\rangle $.

\begin{figure}
[ptb]
\begin{center}
\includegraphics[
height=2.6455in,
width=3.4186in
]%
{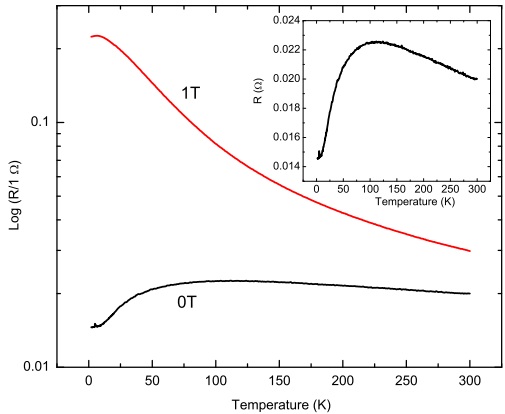}%
\caption{Temperature-dependent resistivity of a typical ZYA sample in zero
field and in a field of 1 T on a logarithmic scale. \ The inset shows the
zero-field data on a linear plot.}%
\label{Fig1}%
\end{center}
\end{figure}

Samples for this study were obtained from ND-MDT Corp. and were of ZYA
quality. Certain samples were subjected to ion implantation by boron at
various dose levels and ion energies ranging from 10 to 40 keV in an effort to induce surface superconductivity within
the ion-implantation range. No evidence for superconductivity was found, and
the MR behavior of irradiated samples, as will be seen, is identical to
as-received material. With one exception, all samples were cut to
approximately 2 mm x 4mm, with four silver contacts placed across the width of
the sample as seen in Fig.\ref{FigM}. The spacing of the inner pair (voltage contacts) was identical for
all samples. Four-terminal resistivity measurements were carried out using a
Quantum Design PPMS system over the temperature range 2K -- 300 K and up to 7
T. The inset to Fig. 1 shows typical data on a ZYA sample which shows
semiconducting-like behavior above $\approx100$ K and metallic behavior
below. As seen in the main portion of Fig.\ref{Fig1}, the low temperature resistance
increases by a factor of 20 in a field of 1 T, and is semiconducting-like
through the entire temperature range.%

\bigskip

\begin{figure}
[ptb]
\begin{center}
\includegraphics[
height=2.6455in,
width=3.4186in
]%
{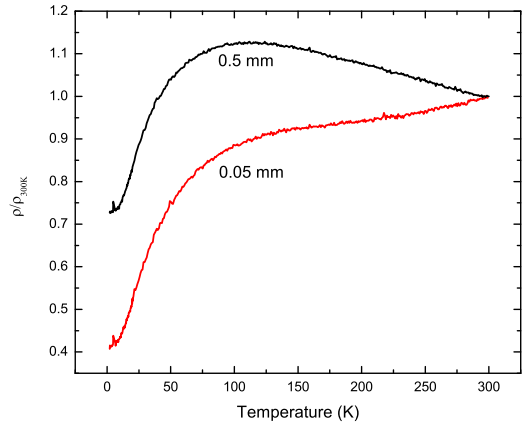}%
\caption{Contrasting zero-field resistance of a 0.5 mm thick sample of ZYA and
that of a thin (0.05 mm) portion of the same sample}%
\label{Fig2}%
\end{center}
\end{figure}

As an example of the wide variation in the temperature dependence of HOPG,
Fig.\ref{Fig2} shows the normalized resistivity a sample of thickness 0.5 mm along
with that of a thin portion of the same sample having a thickness of 0.05 mm.
While the thicker sample shows the same behavior as the sample in Fig.\ref{Fig1}, the
thinner sample is metallic over the entire temperature range.%

\begin{figure}
[ptb]
\begin{center}
\includegraphics[
height=2.6455in,
width=3.4186in
]%
{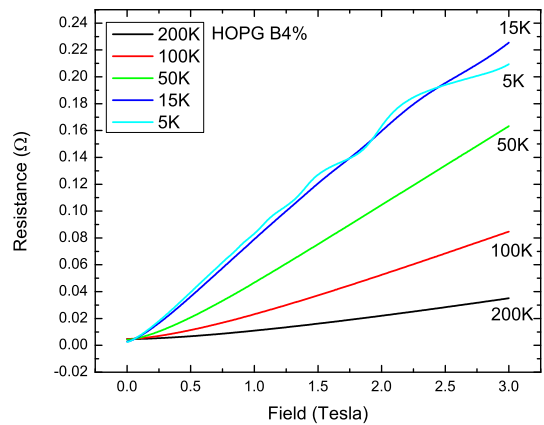}%
\caption{The field dependence of the resistance at five temperatures for a
sample of \ HOPG doped by boron ion-implantation. Note that Shubnikov-de\ Haas
oscillations cause an apparent maximum in R(T) at 3 T, }%
\label{fig3}%
\end{center}
\end{figure}

The nonsaturating MR is shown in Fig.\ref{fig3} for a sample that had been ion
implanted with boron at a level of 3 percent at an energy of 20 keV. At low temperatures, pronounced oscillations are
observed which cause an apparent maximum in the resistivity near 15 K when
plotted versus temperature at a fixed field of 3 T. Such downturns in
resistance at low temperature have been attributed to the existence of
superconducting regions, but are clearly a consequence of quantum
oscillations. The magnetoresistance shows curvature at higher temperatures,
but becomes increasingly linear as the temperature decreases.

\ %

\begin{figure}
[ptb]
\begin{center}
\includegraphics[
height=2.9386in,
width=3.7957in
]%
{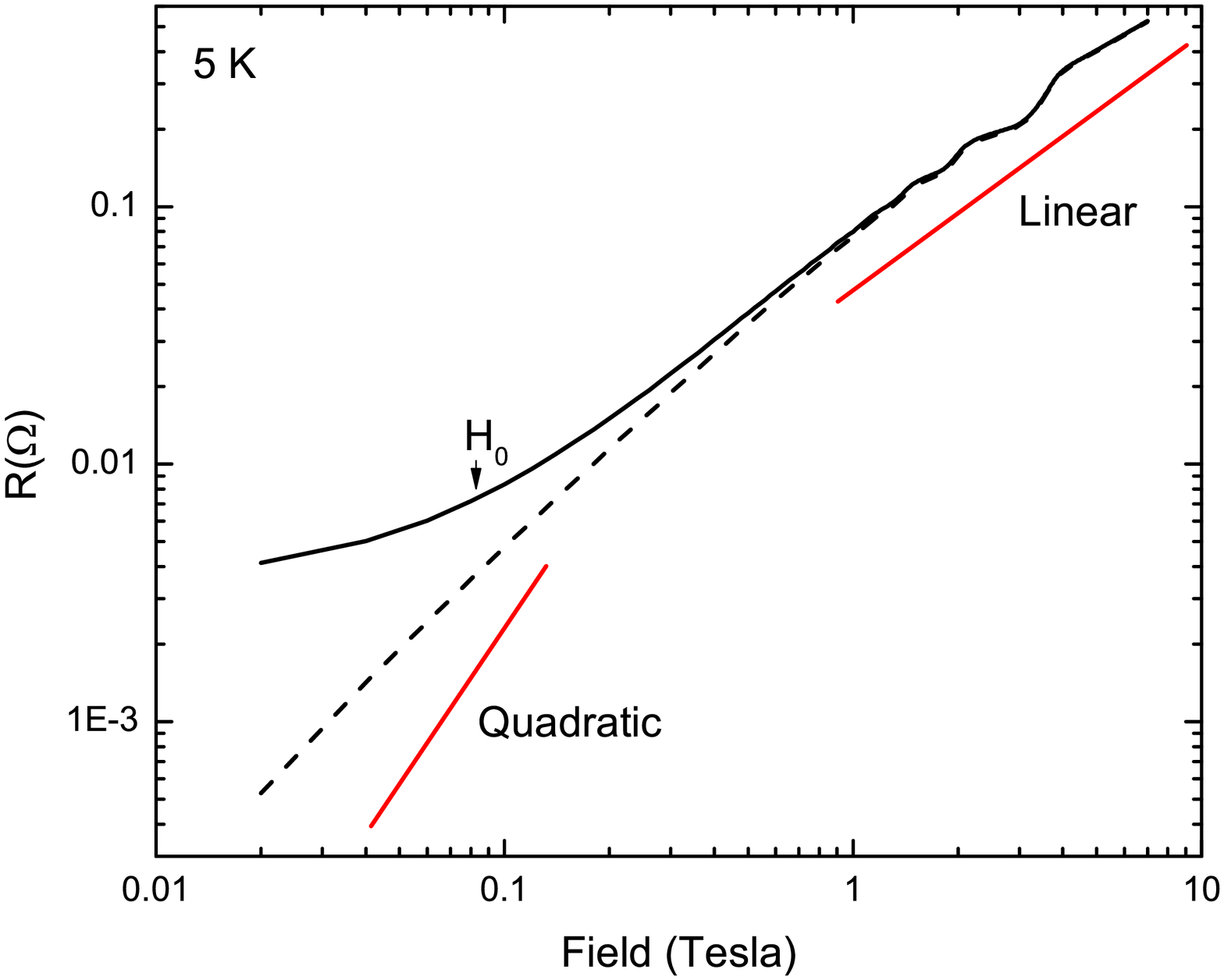}%
\caption{Logarithmic plot of the resistance at 5 K to 7 T. \ The
mgnetoresistance is linear in field (red line) from 0.7 T to 7T, with no
evidence for saturation. \ The dashed curve shows the same data with the
zero-field value subtracted. \ The data are not purely quadratic in field down
to 0.01 T.}%
\label{fig4}%
\end{center}
\end{figure}

\ In Fig.\ref{fig4}, we show the 5K MR of a ZYA sample to 7T. The linear field
dependence is clear. \ The dashed line shows the data with the zero-field
resistance subtracted. \ To the lowest fields measured, the data are not
clearly quadratic in field%

\begin{figure}
[ptb]
\begin{center}
\includegraphics[
height=2.6455in,
width=3.4186in
]%
{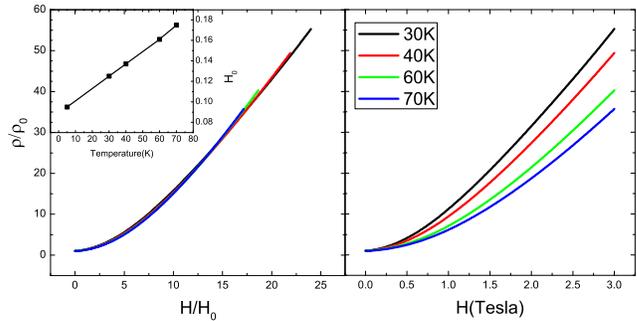}%
\caption{Left figure: resistance curves at four temperatures scaled by a
characteristic field $H_{0}$ that follows the inverse of the mobility,
decreasing with decreasing temperature, with inset showing the dependence of $H_{0}$ on temperature. \ Right figure: the same data before
scaling. }%
\label{fig5}%
\end{center}
\end{figure}

As noted above, Parish and Littlewood\cite{parrish} predict that the crossover
from quadratic to linear field dependence occurs at $H_{0}\approx
1/\left\langle \mu\right\rangle $ when the variation of the mobility among
crystallites is small. At the same time, the MR itself is proposed to scale
with $H_{0}$. We test this prediction in Fig.\ref{fig5}, at temperatures where the MR
oscillations are less evident. $H_{0}$ is chosen for each curve at the point where it deviates from linearity
by a set percentage.  Note that the crossover field $H_{0}$ increases
with temperature, reflecting the decrease in mobility as expected for phonon
scattering in each graphite crystallite. The cross-over field extrapolates to $H_{0}=0.08T$ at
low temperatures.

We turn now to the low temperature magnetoresistance oscillations. Early work
by Williamson et al.\cite{dresselhaus} on the de Haas-van Alphen oscillations
of pyrotylic and single crystal graphite showed their Fermi surface in detail.
The majority electron and hole carriers were found in approximately
ellipsoidal pockets along HKH in the Brillouin zone. There are six such
ellipsoids with one third of their volume within the unit cell. Previous work
on Shubnikov-de Haas oscillations in graphite shows that the majority electron
ellipsoids dominate in the resistivity.\cite{igor}

\begin{figure}
[ptb]
\begin{center}
\includegraphics[
height=2.6455in,
width=3.4186in
]%
{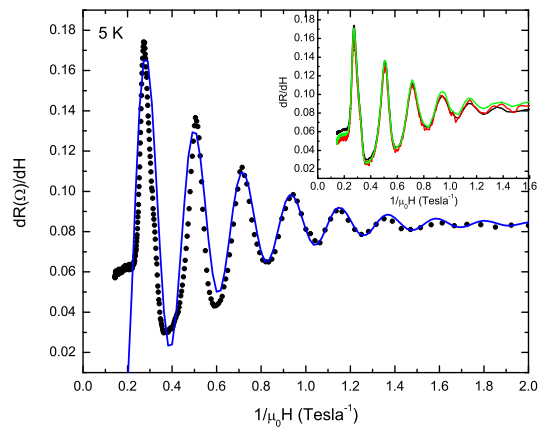}%
\caption{The data for HOPG ZYA (points), shown with the fitting curve (line).
The inset shows the Shubnikov-de Haas oscillations for the ZYA as well as two
ion-implantation doping levels of 3 and 4 percent.}%
\label{fig6}%
\end{center}
\end{figure}

Despite wide variations in temperature dependence of various samples, the MR
is surprisingly similar. The main panel of Fig.\ref{fig6} shows the\ numerical
derivative $dR(H)/dH$ of the data in Fig.\ref{fig4} plotted vs the inverse
field. This method of separating the oscillatory behavior from the linear MR
was reported by Arnt, et al.\cite{amdt} The periodicity in 1/$\mu_{0}H$
identifies these as Shubnikov-de Haas oscillations within each graphite
crystallite. The inset of Fig.\ref{fig6} shows similar curves for both doped
and undoped samples whose zero field resistance can vary by a factor of two.
We note that the samples have the same thickness and contact spacing, so that
the data actually reflects nearly identical magnetoresisistivity, including MR
oscillations.

Fitting the curves in Fig.\ref{fig6} to a version of the SdH oscillations with
a constant relaxation time\cite{ando},%

\begin{equation}
\frac{dR}{d\mu_{0}H}=R^{\prime}+Ae^{-\alpha/\mu_{0}H}\sin(\frac{2\pi\beta}%
{\mu_{0}H}+\phi) \label{Eq1}%
\end{equation}

yields the linear MR coefficient $R\prime=0.084\Omega/T$, oscillation
amplitude $A=0.18\Omega/T$, damping factor $\alpha=2.7T$; period $\beta=4.6T$
and phase factor $\phi=-0.12\pi$. The standard expressions\cite{ando,tan} for
Shubnikov-de Haas oscillations give
\begin{equation}
\beta=\frac{\hbar S_{F}}{2\pi e} \label{eq2}%
\end{equation}

where $S_{F}$ is the cross sectional area of the Fermi surface. We obtain
$k_{F}=(S_{F}/\pi)^{1/2}=1.2\times10^{8}m^{-1}$, which leads us to a 2D
carrier density of%
\begin{equation}
n_{2D}=\frac{k_{F}^{2}}{2\pi}=2\times10^{15}m^{-2} \label{eq3}%
\end{equation}
The Fermi surface for holes can be seen weakly in de Haas-van Alphen
oscillations but not SdH.\cite{dresselhaus}

Studies of microcrystalline graphite\cite{amdt} find $k_{F}\approx
6\times10^{7}$m$^{-1}$ with a 2D carrier density $\approx6\times10^{14}$
m$^{-2}$. That value agrees with our result to within a factor of 2.
We note that each Landau Level contains a 2D density of states of $H/{2\Phi_{0}} ={(H/{4{\rm T}})} \times 10^{15}$ m$^{-2}$, where $\Phi_{0}$ is the flux quantum.  At the first peak in Fig.\ref{fig6}, where $H=$ 4 T, only the first and second Landau levels are filled, each with $10^{15}$ m$^{-2}$.  As the field is increased through 4 T, the second Landau level begins to empty and there are no further oscillations.

Treating the majority electron pocket as an ellipsoid of radius $k_{F}$ with a
height $\frac{1}{2}$ that of the Brillouin zone, we arrive at a 3D carrier
density of $2.3\times10^{24}$m$^{-3}$. This is in reasonable agreement with the
value given by band theory of $3.3\times10^{24}$m$^{-3}$.\cite{dresselhaus}

The damping factor provides a direct measure of the carrier mobility
$\mu$
through $\alpha=\pi/\mu$ or $\mu=1.2\times10^{4}$ cm$^{2}$/Vs. This
corresponds to a mean-free path $\approx0.1$
$\mu$%
m, indicating that the crystallites contributing to the S-dH oscillations and
the large linear MR are substantially defect free. Expressed in field terms,
the mobility is $\mu$ $=$ 1.2 T$^{-1}$. It is also important to remember that this mobility is an average over individual crystallites, and not that of the bulk.  If, as predicted, the cross over
between quadratic and linear MR regions is inversely proportional to $\mu$, we
would expect $H_{0}=$ 0.8 T. In fact, the scaling fields in Fig.\ref{fig5}
extrapolate to a low-temperature value of 0.08 T. In the Parish-Littlewood
model, the crossover can be dominated by the spread in mobility, but this
would require that $\Delta\mu$ $\simeq$12 T$^{-1}$, which does not seem
reasonable. Perhaps further refinement of that model will provide the needed
proportionality factor $\mu H_{0}\simeq0.13$.

In conclusion, while there is a significant variability in the temperature
dependence of the resistance of samples without an applied field, even between
portions taken from the same sample, and particularly for ion-implanted samples, the large non-saturating
magnetoresistance is consistent throughout. This MR is characteristic of the
Parrish-Littlewood model where the currents flow in a nonhomogeneous system of graphite crystallite disks perpendicular to the electric field in the presence of an
applied magnetic field as shown in Fig.\ref{FigM}. \ The numerical simulation in the model predicts that
$\Delta R\approx R(H=0)H/H_{0},$ when the cross over is dominated by the
mobility \ From Eq. (\ref{Eq1}) and Fig \ref{fig4} we have $H_{0}=0.08$ T and
$R(H=0)=4$ m$\Omega$ \ which predicts $\Delta R=0.35$ $\Omega$ at 7\ T, which
compares well with the data in Fig.\ref{fig4} \ We note that saturation is
not predicted to occur in this model until $\mu H\gtrapprox20$; our value at
7\ T is only $\mu H=8.4.$ The SdH oscillations are in agreement with previous
results for the majority-electron Fermi surface and give a majority electron
density consistent with band theory. The fact that only the majority
electrons contribute to the SdH oscillations, allows us to calculate the mean
free path, showing the crystallites comprising the HOPG sample to be
relatively defect free.  It means that the ion-implanted region that has many defects due to the impact of ions does not contribute to the SdH oscillations, implying that the current distribution does spread far below the ion-implanted surface region so that the average mobility is not significantly changed by ion-implantation.  No evidence of ion-implantation induced superconductivity has been detected for the ion energies and does used.


\begin{thebibliography}{99}                                                                                               %


\bibitem {dachs}\textit{Neutron Diffraction}, edited by H. Dachs,
Springer-Verlag, Berlin, 1978, p. 26

\bibitem {baro}A. M. Bar\'{o} et al., Nature \textbf{315}, 253 (1985)

\bibitem {Kopelevich2000}Y. Kopelevich, P. Esquinazi, J. H. S. Torres, and S.
Moehlecke, J. Low Temp. Phys. \textbf{119}, 691 (2000).

\bibitem {Kopelevich1999}Y. Kopelevich, V.V. Lemanov, S. Moehlecke, and J.H.S.
Torres, Physics of the Solid State \textbf{41}, 1959 (1999).

\bibitem {amdt}A. Arndt, et al. Phys. Rev,B \textbf{80}, 195402 (2009)

\bibitem {kempa}H. Kempa, Y. Kopelevich1, F. Mrowka, A. Setzer, J.H.S.
Torres1, R. H\"{o}hne, and P. Esquinazi, Solid State Commun. \textbf{115}, 539 (2000)

\bibitem {Kopelevich2003}Y. Kopelevich, J. H. S. Torres, R. R. da Silva, F.
Mrowka, H. Kempa, and P. Esquinazi, Phys. Rev. Lett. \textbf{90}, 156402 (2003).

\bibitem {hussman}A. Hussmann, J. B. Betts, G. S. Boebinger, A. Migliori, T.
F. Rosenbaum and M.-L. Saboungi, Nature \textbf{417}, 421 (2002).

\bibitem {parrish}M. M. Parish and P. B. Littlewood, Phys. Rev. \textbf{B72,}
094417 (2005).

\bibitem {dresselhaus}S. J. Williamson, S. Foner, and M. S. Dresselhaus, Phys.
Rev. \textbf{140,} 4A (1965).

\bibitem {igor}Igor A. Luk'yanchuk, Y. Kopelevich, Phys. Rev. \textbf{93},
166402 (2004)

\bibitem {ando}T. Ando, J. Phys. Soc. Jpn \textbf{37}, 1233 (1974).

\bibitem {tan}Z. Tan, C. Tan, L. Ma, G. T. Liu, L. Lu and C. L. Yang, Phys
Rev. \textbf{B 84}, 115429 (2011).

\bibitem {zhang}Yuanbo Zhang, Yan-Wen Tan, Horst L. Stormer, and Philip Kim,
Nature \textbf{438}, 201-204 (2005).
\end{thebibliography}
\end{document}